\documentclass{INTERSPEECH2023}
\usepackage{diagbox}

\interspeechcameraready

\title{Multi-Level Knowledge Distillation for Speech Emotion Recognition in Noisy Conditions}
\name{Yang Liu$^{\dag,1}$, Haoqin Sun$^{\dag,1}$, Geng Chen$^{1}$, Qingyue Wang$^{1}$, Zhen Zhao$^{*,1}$, Xugang Lu$^{2}$, Longbiao Wang$^{3}$}
\address{
  $^{1}$Qingdao University of Science and Technology, Qingdao, China\\$^{2}$National Institute of Information and Communications Technology, Kyoto, Japan\\$^{3}$College of Intelligence and Computing, Tianjin University, Tianjin, China}
\email{zzqust@126.com}

\begin{document}

\maketitle
 
\begin{abstract}
Speech emotion recognition (SER) performance deteriorates significantly in the presence of noise, making it challenging to achieve competitive performance in noisy conditions. To this end, we propose a multi-level knowledge distillation (MLKD) method, which aims to transfer the knowledge from a teacher model trained on clean speech to a simpler student model trained on noisy speech. Specifically, we use clean speech features extracted by the wav2vec-2.0 as the learning goal and train the distil wav2vec-2.0 to approximate the feature extraction ability of the original wav2vec-2.0 under noisy conditions. Furthermore, we leverage the multi-level knowledge of the original wav2vec-2.0 to supervise the single-level output of the distil wav2vec-2.0. We evaluate the effectiveness of our proposed method by conducting extensive experiments using five types of noise-contaminated speech on the IEMOCAP dataset, which show promising results compared to state-of-the-art models.
\end{abstract}

\section{Introduction}

\renewcommand{\thefootnote}{}
\footnotetext{$\dag$ These authors contributed equally to the manuscript.}
\footnotetext{* corresponding author.}

With the continuous development of artificial intelligence technology, the field of affective computing is gradually attracting the interest of many researchers. Speech emotion recognition (SER), as a very important branch of affective computing, is penetrating into people's lives, such as improving human-computer interaction \cite{cowie2001emotion}, helping psychological diagnosis \cite{ringeval2018avec} and enhancing customer service \cite{lee2005toward}. Currently, research on SER systems in laboratory environments has yielded good performance \cite{10045019,liu2022multi}. However, in practical applications, SER systems face more complex and demanding conditions, such as noise from different sources. Therefore, a SER system with better robustness is vital for practical applications.

In previous studies, various methods have been developed to address the problem of reducing the effects of environmental noise while retaining valid emotional information, including robust acoustic features extraction \cite{georgogiannis2012speech}, speech enhancement techniques \cite{tawari2010speech} and blind source separation \cite{zhang2014channel}. With the development and widespread application of deep learning, researchers have proposed neural network models with more complex scales in order to improve the accuracy of SER in noisy environment. For example, Wijayasingha et al. \cite{wijayasingha2021robustness} added synthetic noise to the training data and combined the amplitude spectrogram with a modified delay spectrogram to extract spatio-temporal features to reduce the effect of noise using convolutional neural networks (CNN) and attention mechanism. Zhang et al. \cite{zhang2016facing} adopted an autoencoder with long short-term memory (LSTM) for feature enhancement in noisy condition, which significantly improved the accuracy of SER. However, these techniques require complex processing of speech signals before they could be utilized by SER systems for emotion prediction. The additional time overhead associated with the complex processing of the speech signal is very unfriendly to practical applications. 

\begin{figure*}[t]
  \centering
  \includegraphics[width=6.9in]{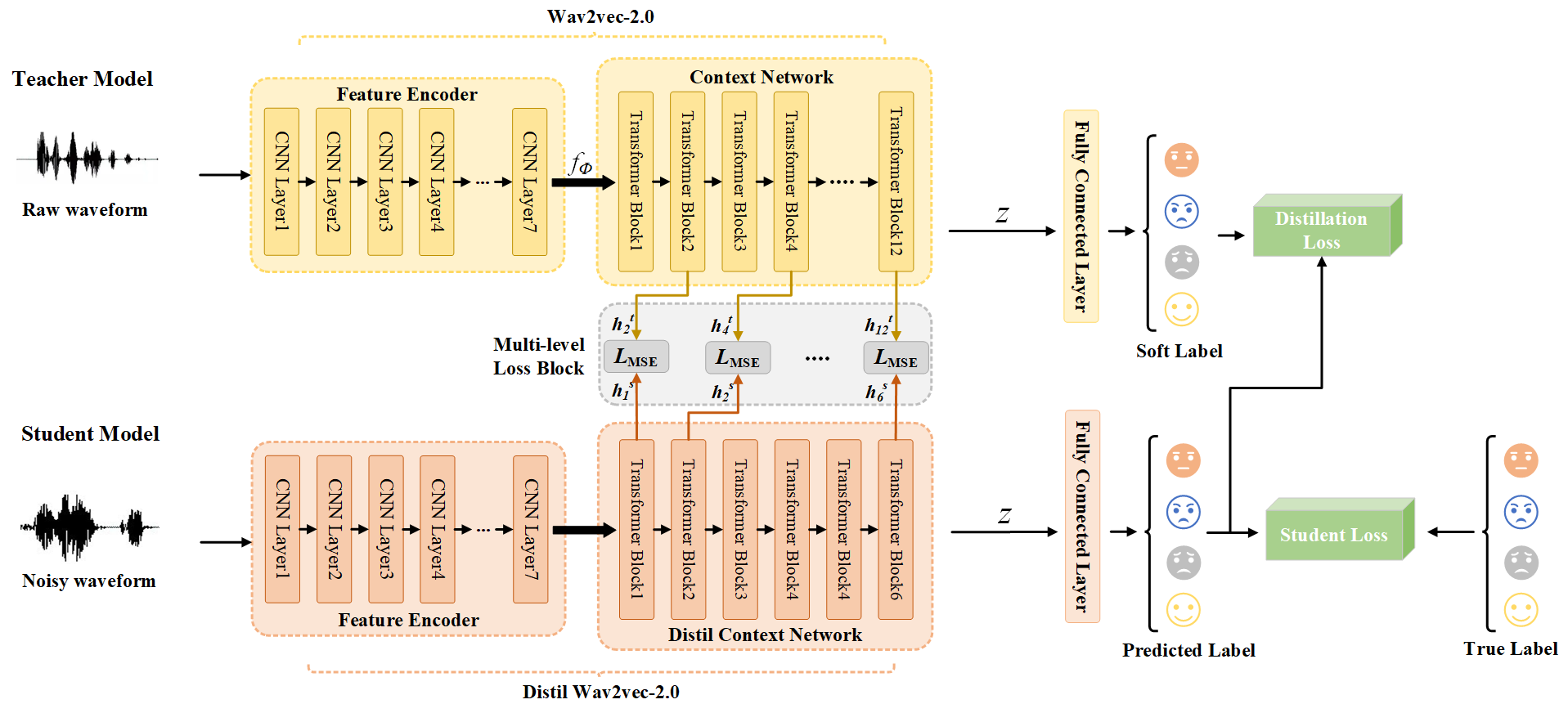}
  \caption{The overall architecture of MLKD, which consists of a teacher model, a student model, and a multi-level loss block. The teacher model consists of seven CNN layers, a pre-trained wav2vec-2.0 and a fully connected layer. The student model consists of a lighter model distil wav2vec-2.0 and a fully-connected layer. The multi-level loss block is used to imitate the feature distribution of the teacher model.} 
  \label{fig:speech_production}
\end{figure*}

In order to solve the above problems, the knowledge distillation (KD) \cite{hinton2015distilling} method may provide a possible solution, which has received a great attention in the field of SER \cite{yun2023end,chen2022electroglottograph}. For example, Tong et al. \cite{tong2022multimodal} proposed a framework combining KD and transfer learning to compress the model and improve the accuracy of emotion recognition with a small amount of data and unbalanced emotion categories. Albanie et al. \cite{albanie2018emotion} and Li et al. \cite{li2021speech} proposed a cross-modal emotion distillation method to transfer the emotion knowledge learned from vision and text modality to speech modality, respectively, which improves the robustness of SER models. In general, KD focuses on training a simpler student model by utilizing additional information in soft labels provided by a teacher model, which also helps the simpler student model still perform well when dealing with emotional speech in noisy conditions. In this paper, we adopt the KD-based architecture to achieve the trade-off between model complexity and performance for SER in noisy conditions. 

This paper proposes a framework based on multi-level knowledge distillation (MLKD) for SER in noisy environments. The training phase of our approach is divided into two steps. In the first step, we train the teacher model in a clean speech corpus. In the second step, we train a distil wav2vec-2.0 as student model with a relatively simple structure in noisy speech corpus. Finally, to ensure that the transferred distilled knowledge has a positive impact on the student model, we use the intermediate layer features of the teacher model to guide the student model in enhancing feature extraction capabilities. Experimental results on the benchmark dataset IEMOCAP show that the absolute gains of our proposed method achieves 18.23\% on unweighted accuracy (UA) on average compared to the baseline system, across all types of noise.  

The main contributions of this study are summarized as follows:

(1) We propose a novel multi-level knowledge distillation framework, aiming to train a student model on noisy speech corpus to approximate the capability of the teacher model trained on clean speech corpus to classify emotions of noisy speech.

(2) We take advantage of the multi-level knowledge of the teacher model to supervise single-level output of the student model, which is a lighter pre-trained distil wav2vec-2.0.

\section{Proposed Method}

The overall structure of the proposed method is shown in Fig. \ref{fig:speech_production}. First, in the teacher model, we use the wav2vec-2.0 containing feature encoder and contextual network to extract the comprehensive emotion feature representation of the raw waveform. Next, after the teacher model is trained, we use the distilled knowledge of the teacher model to provide guidance to the student model on how to obtain better denoising and feature extraction capability in the noisy speech corpus.

\subsection{Multi-Level Knowledge Distillation Framework}

Our proposed MLKD framework consists of a teacher model, a student model, and a multi-level loss block, which aims to enable the structurally simplified student model to imitate the feature extraction ability of the teacher model when trained on a noisy speech corpus. In the MLKD framework, the teacher model is obtained by training on a clean speech corpus and the student model learns features from multiple intermediate layers of the teacher model, which could fully exploit the feature extraction capability of the teacher model, and learn the feature information of the teacher model from shallow to deep. 

The learning process consists of two steps: (1) Model training: we train the teacher model in the clean speech corpus. (2) Distillation: the student model performs two tasks, where the first one is to predict emotion classification, and the second one is to perform a feature regression task using the extracted knowledge in order to simulate the output of the teacher network. We present more details about the knowledge distillation framework below.

\subsubsection{Teacher Model}

\textbf{Pre-trained wav2vec-2.0:} The pre-trained wav2vec-2.0 consists of feature encoder and context network. First, the original audio waveform $x \in R^N$ is encoded by the feature encoder consisting of seven 1D convolutional layers as $f_\phi$, which represents the low-level embedding features of the original waveform. Then, the channel dimension of $f_\phi$ is normalized and then activated by the Gelu function. Next, $f_\phi$ is fed into the context network, which consists of twelve transformer blocks, each containing twelve attention heads. Finally, we take the feature vector $z \in R^{L_b}$ of the last hidden layer of the context network as the input of the fully connected layer, where $L_b$ represents the hidden dimension, which is set to 768. 

\textbf{Fully Connected Layer:} In the final stage of the teacher model, we use a fully connected layer to compress the feature vector $z$ to obtain the soft label $y^t$ for the emotion classification prediction.

\subsubsection{Student Model}

Inspired by the idea of distil-BERT \cite{sanh2019distilbert}, we compress the context network of wav2vec-2.0 from the twelve transformer blocks to six blocks. In the traning process of the student model, first, we add noise with different Signal to Noise Ratios (SNRs) to the clean speech corpus as the input of the student model. Then, the wav2vec-2.0 feature encoder encodes the noisy audio waveforms as low-level embedding features. Next, we initialize the transformer block in student model with the even layer weights of the context network of the teacher model. Finally, the soft prediction $y^s$ is obtained by the fully connected layer.

\subsubsection{Multi-level Loss Block}

We compute the loss of the six transformer blocks so that the student model could learn more feature knowledge from the hidden states of multiple intermediate layers of the teacher model. We then use the transformer blocks of the even layers of the context network of teacher model to guide the student model. The intermediate layer of the teacher model is denoted as $I_{t,i}$, and the intermediate layer of the student model is denoted as $I_{s,i}$, where $I_{t,i}$ = \{2, 4, 6, 8, 10, 12\}, and $I_{s,i}$ = \{1,2,3,4,5,6\}. For the $i$-th intermediate layer, $h_i^{t}$ and $h_i^{s}$ represents the hidden state of the teacher model at the $I_{t,i}$-th layer and student model at the $I_{s,i}$-th layer, respectively. We use the mean square error (MSE) to measure the distance between distillation features. The multi-level loss function is calculated as follow.
\begin{align}
L_\text{MSE}=\frac{1}{N} \sum_{i=0}^{n} (h_i^{t}-h_i^{s})^2
\end{align}

\subsubsection{Training objective}

In the final stage of the distillation framework, we follow the previous work \cite{li2014learning} to transfer knowledge of the decision level of the teacher model to the student model by minimizing the distance between the soft label and soft prediction. Specifically, we use Kullback-Leibler divergence to measure the distance of the emotional probability distribution between the soft label and soft prediction. The distillation loss $L_\text{KL}$ is calculated as follows.
\begin{align}
p_{i}^{t}&=softmax\left(y_{i}^{t} / T\right)\\
p_{i}^{s}&=softmax\left(y_{i}^{s} / T\right)\\
L_{\mathrm{KL}}&=\frac{1}{N} \sum_{i=1}^{n} K L\left(p_{i}^{t} || p_{i}^{s}\right)
\end{align}
where, $p_{i}^{t}$ represents the soft label for the $i$-th sample generated by the teacher model. $p_{i}^{s}$ represents the soft prediction for the $i$-th sample generated by the student model. $T$ represents the distillation temperature coefficient, which is used to soften the probability distribution of the soft label. 

Although the knowledge of the teacher model far exceeds that of the student model, it still has the potential to make errors. If the student model could correctly recognize emotions in addition to the teacher's guidance, it could effectively reduce the possibility of occasional errors by the teacher model. Therefore, the final objective function $L$ is obtained by weighting the distillation loss and the student loss. 
\begin{align}
L_{C}&=-\frac{1}{N} \sum_{i=0}^{N-1} \sum_{k=0}^{K-1} y_{i, k} \ln p_{i, k}\\
L&=\alpha \times L_\text{KL}+(1-\alpha) \times L_{C} +  L_\text{MSE}
\end{align}
where $L_C$ represents cross-entropy loss function, which is defined as student loss. $y_{i, k}$ represents the true probability distribution and the true label of the $i$-th sample as $k$. $p_{i, k}$ represents the probability output of the softmax layer and the probability that the $i$-th sample is predicted to be the $k$-th label.

\section{Experiments and Evaluation Results}
\subsection{Emotion Dataset}

\textbf{IEMOCAP:} The IEMOCAP corpus \cite{busso2008iemocap} contains 10,039 utterances with a sampling rate of 16 kHz. Each utterance in the corpus is annotated with an emotion. Four emotions are selected: happiness (combined with the excitement class), anger, neutral and sadness, for a total of 5531 utterances in our experiments.

\textbf{Noisex-92 database:} The Noisex-92 database \cite{varga1992noisex} is noise data measured in the field by the Speech Research Unit of the Institute of Perceptual Technology in the UK, where all noise files are of 235 seconds duration, using a sampling rate of 19.98 KHz. It consists of 15 categories of noise. We have used 5 different types of noise (voice: babble, F-16 fighter-jet, factory, HF radio and volvo 340) in our experiments.

\subsection{Experimental Setup}

We choose both the distillation temperature $T$ and the weight of distillation loss $\alpha$ by the grid search method and obtain the best performance of the student model when $T$ and $\alpha$ are set to 7 and 0.7, respectively. The pre-trained wav2vec-2.0 is used in our experiments, which is implemented based on the Huggingface transformers repository \cite{wolf2019huggingface}. We use 10-speaker-independent Leave-One-Speaker-Out cross-validation strategy. We choose the AdamW optimizer and the learning rate is $5\times10^{-5}$. We set the epoch to 200 to train the teacher and student models, and the batch size is 8. 

In the experiment, we add noise to the clean signal using the FANT toolkit \cite{hirsch2005fant} at levels of SNRs of 0 dB, 5 dB, 10 dB, 15 dB, and 20 dB. We randomly select a noise to add to each clean speech. Voice activity detector + Non-negative Matrix Factorization (VAD + NMF) \cite{pandharipande2018robust} is chosen as the baseline to compare with our proposed method. UA is chosen to evaluate the performance of the proposed method. 

\begin{table*}[t]
\caption{Emotion recognition accuracy (UA in \%) of the teacher model, the student model trained under supervision and student model trained without supervision on IEMOCAP corpus contaminated by factory noise with six SNR levels.}
\label{tab:example1}
  \centering
\begin{tabular}{l|c|c|c|c|c|c}
\hline
Models       &SNR = 0 dB&SNR = 5 dB&SNR = 10 dB &SNR = 15 dB   & SNR = 20 dB  & clean speech  \\ \hline
Teacher  &41.03 &53.95 &65.00&71.55& 75.75  & 77.78    \\\hline

Student without supervision & 57.88 &61.68 &65.13&67.78&68.00  & 72.18    \\\hline
Student with supervision  & 63.98 & 66.10 &68.50&69.10& 69.57   & 73.86    \\ \hline
\end{tabular}
\end{table*}

\begin{table*}[t]
\caption{Emotion recognition accuracy (UA in \%) on IEMOCAP corpus (5 types of noise with 5 SNR levels). Baseline: VAD + NMF. MLKD: our proposed method.}
\label{tab:example2}
  \centering
\begin{tabular}{c|c|c|c|c|c|c|c|c|c|c}
\hline
                                                                          
Noisy types& \multicolumn{2}{c}{Babble} & \multicolumn{2}{|c}{F16} & \multicolumn{2}{|c}{Factory} & \multicolumn{2}{|c}{HF-channel} & \multicolumn{2}{|c}{Volvo} \\\hline
 \diagbox{SNRs}{Method}&baseline&MLKD&baseline&MLKD&baseline&MLKD&baseline&MLKD&baseline&MLKD  \\\hline
  0 dB & 45.12 & 61.25 & 40.32 & 61.00 & 42.71 & 63.98 & 45.01 & 59.50 & 48.93 & 68.05  \\\hline
  5 dB & 46.87 & 66.90 & 41.03 & 64.07 & 43.93 & 66.10 & 46.73 & 64.65 & 51.87 & 69.18  \\\hline
 10 dB & 53.01 & 68.67 & 44.65 & 68.40 & 46.01 & 68.50 & 48.12 & 67.23 & 54.76 & 69.55  \\\hline
 15 dB & 55.81 & 69.40 & 48.11 & 69.15 & 48.12 & 69.10 & 48.76 & 70.35 & 56.95 & 70.25  \\\hline
 20 dB & 56.12 & 71.70 & 53.51 & 70.62 & 49.13 & 69.57 & 50.98 & 70.60 & 59.10 & 70.75  \\\hline
  \multicolumn{1}{c|}{Mean} & \multicolumn{1}{c}{53.01} & \multicolumn{1}{c}{\textbf{67.58}} & \multicolumn{1}{c}{44.65} & \multicolumn{1}{c}{\textbf{66.64}} & \multicolumn{1}{c}{46.01} & \multicolumn{1}{c}{\textbf{67.45}} & \multicolumn{1}{c}{48.12} & \multicolumn{1}{c}{\textbf{66.46}} & \multicolumn{1}{c}{54.76} & \multicolumn{1}{c}{\textbf{69.55}} \\\hline
\end{tabular}
\end{table*}

\begin{table}[th]
  \caption{Comparison of the number of parameters and inference time between the teacher and student models for a batch size of 507.}
  \label{tab:example3}
  \centering
  \begin{tabular}{ccc}
    \toprule
    Models   & Param       & Time \\ 
    \midrule
   Teacher   & 360.17      & 926  \\
   Student   & 197.91      & 444  \\ 
    \bottomrule
  \end{tabular}
\end{table}

\subsection{Evaluation}

To evaluate the effectiveness of the proposed MLKD framework for SER, two different approaches are implemented to train the student model. The first approach is to train the student model independently, while the second approach is to train it under the supervision of the teacher model. After training, the student model is tested under factory noise conditions with six SNR levels. The implementation results show that the student model under the supervision of the teacher model ourperforms the student model trained independently. The absolute improvement of UA is 6.10\% (0 dB), 4.42\% (5 dB), 3.37\% (10 dB), 1.32\% (15 dB), 1.57\% (20 dB), and 1.68\% (clean speech), with an average improvement of 3.08\% across all types of noise. The teacher model focuses on how to extract comprehensive emotion information in the clean speech. The student model could learn the knowledge of emotion feature representations from the teacher model and apply it to the noisy speech, thus alleviating the effect of noise on emotion recognition. It should be noted that when the SNR is above 15 dB, the contaminated speech is close to clean speech, and in this case, the recognition performance of the student model is lower than that of the teacher model. However, the performance of student model gradually surpasses that of teacher model as the SNR decreases. Therefore, our proposed method performs better at low SNRs, while it is slightly inferior to the teacher model at high SNRs. These findings suggest that our proposed MLKD framework could effectively improve the performance of the student model, which is more suitable for SER in noisy conditions.

To evaluate the computation complexity of the proposed student model, we compare the number of parameters and reasoning time between the teacher model and the student model, as presented in Table~\ref{tab:example3}. The teacher model occupies 360.17 MB of memory space, whereas the student model occupies only 197 MB of memory space, indicating that the proposed student model has 45\% fewer parameters compared to the teacher model. Furthermore, the inference time of both models is compared on the same epoch using an Intel(R) Xeon(R) CPU Silver 4110 of 2.10 GHz 64-bit processor with 16 GB RAM. The teacher model takes 926 seconds to recognize 507 samples, while the student model takes only 444 seconds. The results show that in the case of clean speech, the proposed student model achieves 52\% less inference time with a sacrifice of 5\% performance compared to the teacher model. Therefore, we conclude that the proposed student model is lighter and faster while maintaining acceptable performance in practical applications.

\subsection{Comparison with State-of-the-art Approaches}

To demonstrate the effectiveness of our proposed MLKD framework for SER, we conducted experiments to compare it with a baseline system (VAD + NMF). The results are showed in Table~\ref{tab:example2}, which shows the emotion recognition accuracy of our proposed method under five types of noise conditions (Babble, F-16, Factory, HF-channel, Volvo) at five different SNR levels (0 dB, 5 dB, 10 dB, 15 dB, 20 dB).

For all SNR levels of each type of noise, the average absolute improvement on UA using our proposed method over the baseline is 14.57\% (Babble), 21.99\% (F-16), 21.44\% (Factory), 18.34\% (HF-channel), 14.79\% (Volvo) (with an average improvement of 18.23\% across all types of noise). Specifically, when the SNR is reduced from 20 dB to 15 dB, the UA of our supervised student model decreases by only 1.00\%, while the UA of the baseline decreases by 2.22\%. Although the performance of both models decreases, the degree of decrease of our student model is less significant, indicating the superiority of our student model. 

Furthermore, when the SNR is reduced to 10 dB, the performance of both methods continues to decrease. However, the performance of our supervised student model is the least affected, where the UA remains close to 68\%. In addition, the advantage of our proposed method is more evident at SNRs of 0 dB and 5 dB. The baseline method shows an average performance degradation of about 8\%, while the performance of our proposed method only decreases by 4\%. Therefore, our student model is more robust and has better generalization ability. Moreover, the proposed MLKD approach enables the student model to extract clean emotion features from the corpus in noisy environments, which allows our student model to maintain impressive performance in the noisy speech corpus. Overall, the experimental results show that our proposed method outperforms the baseline system, and the student model trained with MLKD is more suitable for SER in noisy environments.

\section{Conclusions}

In this paper, we present a multi-level knowledge distillation framework that includes a teacher model, a student model, and a multi-level loss block. Our framework aims to transfer the feature extraction ability of the teacher model in the clean speech corpus to a more lightweight student model trained with noisy speech corpus. We first use the wav2vec-2.0 model to extract comprehensive acoustic features. Next, we design a distilled student model, distil wav2vec-2.0, where we utilize the transformer block weights of the even layer of the teacher model to initialize the student model to enhance its training and convergence. Lastly, we use the multi-level loss block, which uses multiple intermediate layer representations of the teacher model to guide the output of intermediate layers of the student model. Our series of experimental results show that our student model maintains a relatively impressive ability for emotion recognition of contaminated speech under the guidance of the teacher model. Compared to the state-of-the-art approach, the absolute gains of our proposed method achieve an average of 18.23\% on UA across all types of noise.

\section{Acknowledgments}
This work is supported by National Natural Science Foundation of China (NSFC) (62201314, 62201571), Natural Science Foundation of Shandong Province (No.ZR2020QF007). Key Technology Tackling and Industrialization Demonstration Projects of Qingdao (23-1-2-qdjh-18-gx)
% ~\\
% ~\\
% ~\\
% ~\\
% ~\\

\bibliographystyle{IEEEtran}
\bibliography{mybib}

\end{document}